\documentclass[12pt]{article}
\usepackage{graphics}

\newcommand\bmat{\left( \begin{array}{cc}}
\newcommand\emat{\end{array}\right)}

\newcommand{\sW}{\sin\theta_W}

\def\msbar{\ifmmode{\overline{\rm MS}} \else{$\overline{\rm MS}$} \fi}
\def\drbar{\ifmmode{\overline{\rm DR}} \else{$\overline{\rm DR}$} \fi}

\def\ti              {\tilde}

\def\a               {\alpha}
\def\b               {\beta}
\def\d               {\delta}
\def\D               {\Delta}
\def\e               {\epsilon}

\def\G               {\Gamma}

\def\t               {\theta}
\def\s               {\sigma}
\def\S               {\Sigma}
\def\x               {\chi}
\def\f               {\phi}

\def\sq              {{\ti q}}

\def\sf              {{\ti f}}
\def\sfe              {{\ti f}}

\def\st              {{\ti t}}
\def\sb              {{\ti b}}

\def\ch                {{\ti \chi}}
\newcommand{\cha}[1]   {{\ti \x^+_{#1}}}
\newcommand{\cham}[1]  {{\ti \x^-_{#1}}}
\newcommand{\chapm}[1]  {{\ti \x^\pm_{#1}}}
\newcommand{\neu}[1]   {{\ti \x^0_{#1}}}

\def\xp                  {{chargino }}
\def\xz                  {{neutralino }}
\def\ren                {{renormalize }}

\newcommand{\mcha}[1]   {m_{\ti \x^+_{#1}}}

\newcommand{\mnt}[1]   {m_{\ti \x^0_{#1}}}

\def\sg              {{\ti g}}

\def\tb              {\tan\beta}
\def\tsf              {\theta_\sfe}

\newcommand{\bea}{\begin{eqnarray}}
\newcommand{\eea}{\end{eqnarray}}
\newcommand{\beq}{\begin{equation}}
\newcommand{\eeq}{\end{equation}}

\newcommand\mcps{{m_j^2}}         
\newcommand\mcss{{m_i^2}}         
\newcommand\mcp{{m_j}}            
\newcommand\mcs{{m_i}}            

\def\half             {{\frac12}}

\def\non             {\nonumber}

\newcommand\ksla{{k \hspace{-2.0mm} \slash}}

\newcommand\psla{{p \hspace{-1.8mm} \slash}}

\newcommand{\LL}{{\cal{L}}}
\newcommand{\MM}{{\cal{M}}}

\renewcommand\d{\delta}

\begin{document}

\pagestyle{empty} \vspace*{-1cm}

\begin{flushright}
  HEPHY-PUB 762/02 \\
  hep-ph/0209137
\end{flushright}

\vspace*{2cm}

\begin{center}
\begin{Large} \bf
Radiative Corrections to SUSY processes\\[2mm] in the MSSM
\end{Large}

\vspace{10mm}

{\large W.~Majerotto
\footnote{Plenary talk given at the SUSY02
Conference, June 17--23, 2002, DESY, Hamburg.}
}

\vspace{6mm}

\begin{tabular}{l}
 {\it Institut f\"ur Hochenergiephysik der \"Osterreichischen
 Akademie der Wissenschaften,}\\
 {\it A--1050 Vienna, Austria}
\end{tabular}

\vspace{20mm}

\begin{abstract}
We review the status of radiative corrections to SUSY processes.
We present the method of the on--shell renormalization for the sfermion and
the chargino/neutralino system and work out the appropriate
renormalization conditions. In particular, we discuss slepton, squark,
chargino and neutralino production in $e^+ e^-$ collisions and the
two--body decays of sfermions and of Higgs
bosons into SUSY particles. It is necessary to
take into account radiative corrections in the precision studies
possible at a future linear $e^+ e^-$ collider.
\end{abstract}
\end{center}

\vfill
\newpage
\pagestyle{plain} \setcounter{page}{2}

\section{Introduction \label{sec:intro}}

Why should we bother about radiative corrections to processes with
supersymmetric (SUSY) particles when no SUSY particle has been
found yet? Clearly, one expects that the next generation of high
energy physics experiments at Tevatron, LHC, and at a future linear
$e^+e^-$ collider will discover supersymmetric particles. But it
is also obvious that precision studies are needed in order to
single out the right supersymmetric model and to reconstruct its
fundamental parameters. In particular, at a linear $e^+e^-$
collider it will be possible to perform  measurements with
high precision. For instance, at TESLA \cite{TESLA} the precision
of the mass determination of charginos or neutralinos will be
$\Delta m_{\ti{\x}^{\pm,0}}=0.1-1$~GeV and of sleptons
(sneutrinos) $\Delta m_{\tilde{l},\tilde{\nu}}=0.05-3$~GeV. To
match this accuracy, it is inevitable to include higher order
corrections. Moreover, in some cases these corrections can be very
large.

In the following we will discuss SUSY particle
production in $e^+e^-$~colliders and SUSY particle decays within the
Minimal Supersymmetric Standard Model (MSSM):
 \bea
 e^+e^- &\rightarrow& \ti\x^+_i\ti\x^-_j
 \hspace{30mm}
 i,j=1,2 \label{eq:ee2xx}\\
 &\rightarrow& \ti\x^0_l\ti\x^0_k
 \hspace{31.3mm}
 l,k=1,\ldots,4
 \non \\
 e^+e^- &\rightarrow& \sfe_i \ \bar{\!\!\tilde f}_{\!\!j}
 \hspace{15.6mm} {\rm with} \hspace{13mm}
 f=l,q\;\;(q=b,t)
 \label{eq:ee2sfsf}\\
  e^+e^- &\rightarrow& q\ \bar{\!\ti{q}}\,\ti{g}\,(g)
  \hspace{9mm} {\rm with\; light\; quarks}
  \label{eq:ee2qqg}
\eea
 and the decays
\bea
 \sfe_{1,2} &\rightarrow& f\,\ti{\x}^\pm_i\non\\
&\rightarrow& f\,\ti{\x}^0_l
 \label{eq:sf2fx}\\
 \sfe_i &\rightarrow& \sfe_j+(W,Z,H_m)
 \hspace{19.4mm}
 m=1,\ldots,4
 \label{eq:sf2sfWZH}\\
 H_m &\rightarrow& \ti{\x}_i^{+}\ti{\x}_j^{-},\;\ti{\x}_l^{0}\ti{\x}_k^{0}
 \hspace{11mm}{\rm or}\hspace{11mm}
 \ti{\x}^{+,\,0}_{i,\,k} \rightarrow \ti{\x}^{+,\,0}_{j,\,l} + H_m
 \label{eq:H2xx}\\
 H_m &\rightarrow& \ti{q}_i \ \bar{\!\ti{q}}_{\!j}
 \label{eq:H2qq}
\eea

with $H_m=\{h^0,H^0,A^0,H^\pm\}$. $\sfe_i$ (with
$i=1,2$) are the mass eigenstates of the sfermions.

For calculating higher order corrections, one has to employ
appropriate renormalization conditions, or equivalently,
one has to fix the counter terms for the SUSY parameters. We will
discuss these fixings in the {\underline{on--shell}} scheme. This
method should of course preserve the symmetries (supersymmetry and
gauge symmetry) and, if possible, should be process independent and
lead to numerically stable results. We will take in the following
the point of view of a practitioner, referring  for a general
discussion on the renormalization of the MSSM using an algebraic
method to~\cite{Hollik1}.

\section{General Method \label{sec:general}}

We start from the Lagrangian of the MSSM with its gauge--fixing and
ghost part, without writing it explicitly:
\beq
 \LL=\LL_{\rm MSSM}+\LL_{\rm gauge\;fixing}+\LL_{\rm ghost}.
 \label{eq:LagrMSSM}
\eeq
We get the renormalized Lagrangian by transforming all fields by\\
$\f_i\rightarrow\sqrt{Z_i}\f_i=(1+\half\d{Z_i})\f_i$ and all
parameters, such as couplings, by $c_k\rightarrow c_k+\d c_k$:
\beq
 \LL^{\rm ren}=\LL-\d\LL,
 \label{eq:LagrRen}
\eeq where $\LL^{\rm ren}$ has the same form as $\LL$, and $\d\LL$
is the counter term part which renders $\LL^{\rm ren}$ finite.
$\d\LL$ contains all counter terms to the parameters which have to
be fixed appropriately. We will use here the \drbar (dimensional
reduction) scheme~\cite{drbarscheme} with the dimension $n=4-r\e$,
$r=0$. It conserves supersymmetry at least up to one--loop. As
already mentioned, we employ the on--shell scheme with the particle
masses as pole masses and the parameters determined on--shell. As a
consequence, there is no scale dependence. We also use the $R_\xi$
gauge~\cite{rxigauge}.

\subsection{Mixing angle of sfermions \label{subsec:sfermixangle}}

The SUSY partners $\sfe_L$, $\sfe_R$ to the fermion $f$ mix with
each other due to the SU(2)$\times$U(1) breaking. The mass
eigenstates $\sfe_i$, $i=1,2$ are
\bea
  \sfe_i=R\sfe_\a,
  \hspace{30mm}
  \a=L,R
  \non
\eea
with
$R=\left(
\begin{array}{cc}\cos{\tsf}&\sin{\tsf}\\-\sin{\tsf}&\cos{\tsf}
\end{array}
\right)$.
The mixing angle $\tsf$ is a measurable quantity.
\\
The corresponding potential $V$ at tree--level is
\bea
 \non
 V&=&\left(\sfe^*_L,\sfe^*_R\right)
 \underbrace{\left(\begin{array}{cc}m_{LL}^2&m_{LR}^2\\m_{RL}^2&m_{RR}^2
\end{array}\right)}_\MM
 \left(\begin{array}{c}\sfe_L\\\sfe_R
 \end{array}\right)\,+\,{\rm h.c.}\\
 &=&\sfe_i\left(\MM_D\right)_{ii}\sfe_i,
 \hspace{20mm}
 \MM_D=R\,\MM\,R^\dagger=
 \left(\begin{array}{cc}m_{\sfe_1}^2&0\\0&m_{\sfe_2}^2\end{array}\right)
\non
\eea
It is renormalized by
\bea
  \sfe_i^0&=&\left(1+\half{Z}\right)\,\sfe_i \non\\
  R^0&=&R+\d{R}=\left(1+\d{r}\right)R \non\\[1mm]
  \MM^0&=&\MM+\d\MM \non
\eea
First we observe that due to the unitarity of $R^0$ and $R$, $\d{r}$
is antihermitian, $\d{r}=-\d{r}^\dagger$. On the other hand, one
can decompose the wave--function renormalization counter term into a
hermitian and antihermitian part:
\bea
 \d{Z}=\half\left(\d{Z}+\d{Z}^\dagger\right)
 +\half\left(\d{Z}-\d{Z}^\dagger\right)\,,
 \non
\eea
It is therefore natural to fix $\d{r}$ such that it cancels the
antihermitian part of $\d{Z}$, i.e.
\bea
  \non
  \d{R}&=&\frac14\left[\d{Z}^\sfe-\left(\d{Z}^\sfe\right)^\dagger\right]R\,,
  \\[2mm]
  \d\tsf&=&\frac14\left(\d{Z}^\sf_{12}-\d{Z}^\sf_{21}\right)
  ~=~
  \frac
  {\S_{12}\big(m_{\sf_2}^2\big)+\S_{21}\big(m_{\sf_1}^2\big)}
  {2\Big(m_{\sf_1}^2-m_{\sf_2}^2\Big)} \,.
\eea 
Here $\S_{12}$ are the non-diagonal self--energies of
$\sf_i$.\\ This is a process independent fixing nowadays
conventionally used \cite{guasch1, 21a}. This fixing is, however, in
general gauge dependent. It was shown in~\cite{youichigauge} that
this gauge dependence can be avoided or, equivalently the result
in the $\xi=1$ gauge can be regarded as the gauge independent one.

The mixing angle $\a$ in the $h^0$--$H^0$ system can be treated in a
similar way as above, leading to the counter term
\beq
 \d\a=\frac14\left(\d{Z}_{21}-\d{Z}_{12}\right)\,,
 \label{eq:dalpha}
\eeq
where the index $1$ is for $h^0$ and $2$ for $H^0$.

\subsection{Renormalization of the chargino mass matrix
\label{subsec:chargmass}}

Here we closely follow the method outlined in~\cite{chmasscorr}.\\
In the MSSM the chargino mass at tree--level is given by
\begin{equation}
        X = \bmat M & \sqrt{2}\, m_W \sin\beta \\
                 \sqrt{2}\, m_W \cos\beta & \mu \emat \, .
  \label{eq:chargmat}
\end{equation}
It is diagonalized by the two real $(2 \times 2)$ matrices $U$ and
$V$:
\begin{equation}
        U\, X\, V^T = M_D = \bmat
m_{\tilde\chi^+_1} & 0 \\ 0 & m_{\tilde\chi^+_2} \emat \, ,
 \label{eq:UXV}
\end{equation}
with $\mcha{1}$ and $\mcha{2}$ the physical masses of the
charginos (choosing $\mcha{1}<\mcha{2}$) where
$$
  \x^L_i=V_{ij}\,\psi^L_j\,,
  \hspace{15mm}
  \x^R_i=U_{ij}\,\psi^R_j\,,
  \hspace{25mm}
  i=1,2\,,
$$
forming the Dirac spinor
$$
  \ch^+_i=\left(\begin{array}{c}\x_i^L\\\bar{\x}_i^R\end{array}\right)\,.
$$
Then we renormalize by performing the shifts
\bea
\non
X^0 & = &X + \delta X\, , \\
\non
U^0 & = & U + \delta U\, , \\
\non
V^0 & = & V + \delta V\, , \\
\psi_{L,R}^0 & = & \left(1 + \frac{1}{2}\, \delta Z^{\ch^+}_{L,R}\right)
\,\psi_{L,R}\;\; .\non
\eea
$Z^{\ch^+}_{L,R}$ are the wave--function renormalization
constants. Proceeding as before and demanding that the counter
terms $\d{U}$ and $\d{V}$ cancel the antisymmetric parts of the
wave--function corrections, we get the fixing conditions
\begin{eqnarray}
\non
 \delta U & = & \frac{1}{4}\,(\delta
Z_R^{\tilde\chi^+}
 - \delta Z_R^{\tilde\chi^+\,T})\, U\,,\\
 \label{eq:dUdV}
\delta V & = & \frac{1}{4}\,(\delta Z_L^{\tilde\chi^+}
 - \delta Z_L^{\tilde\chi^+\,T})\, V\, .
\end{eqnarray}
 $Z^{\ch^+}_{L,R}$ are given by ($i\neq j$):
 \bea
 \left(\d{Z}^{\ch^+}_L\right)_{ij}
 &=&
 \frac{2}{\mcss-\mcps}\,
 {\rm Re}\Bigg\{
 \Pi^L_{ij}(\mcps)\,\mcps
 \,+\,
  \Pi^R_{ij}(\mcps)\,\mcp\,\mcs
 \non \\&&\hspace{40mm}
 \,+\,
  \Pi^{S,L}_{ij}(\mcps)\,\mcs
 \,+\,
 \Pi^{S,R}_{ij}(\mcps)\,\mcp
 \Bigg\}\,.
 \label{eq:dZL}
 \eea
$\left(\d{Z}^{\ch^+}_R\right)_{ij}$ is obtained by replacing $L\leftrightarrow
R$ in Eq.~(\ref{eq:dZL}). Here $m_i=m_{\ch^+_i}$. The mass shifts
$\d m_i$ are given by
 \beq
 \d\mcs\;=\;
 \frac12\,{\rm Re}\Bigg\{
 \mcs\bigg[
  \Pi^L_{ii}(\mcss)+ \Pi^R_{ii}(\mcss)
  \bigg]+
 \Pi^{S,L}_{ii}(\mcss)+\Pi^{S,R}_{ii}(\mcss)
 \Bigg\}\,.
 \label{eq:dmcharg}
 \eeq
The $\Pi_{ij}$'s are self--energies according to the decomposition
of the two--point function of the chargino $\ch^+_i$ and
$\ch^+_j$
\bea
 i\,\hat{\Gamma}_{ij}(k)
 =
 i\,\d_{ij}\left(\ksla-m_{j}\right)+
 i\,\ksla\bigg[P_L\hat{\Pi}^L_{ij}(k^2)+P_R\hat{\Pi}^R_{ij}(k^2)\bigg]
 +
 i\,\hat{\Pi}^{S,L}_{ij}(k^2)P_L +
 i\,\hat{\Pi}^{S,R}_{ij}(k^2)P_R
 \,,
 \label{eq:hatGamma}
 \eea
 where the hat denotes the renormalized quantities. Finally, the
 shift of the mass matrix $\d{X}$ follows from
 $\d{X}=\d\left(U^T\MM_D V\right)$:
\bea
\label{eq:dX}
\left(\d{X}_{ij}\right)=\half\sum_{k,l=1}^2
U_{ki}V_{lj}\;{\rm Re}
\Big[\Pi^L_{kl}(m_k^2)\,m_k + \Pi^R_{kl}(m_l^2)\,m_l+
 \Pi^{S,L}_{kl}(m_k^2)+\Pi^{S,R}_{lk}(m_l^2)\Big]\,,
\eea
with $\Pi^{S,L}_{kl}=\Pi^{S,R}_{lk}$. \\[0mm]

{\bf\boldmath Renormalization of $M$ and $\mu$}: In principle, one
can fix $M$ and $\mu$ by the chargino or the neutralino sector. We
choose the chargino sector, that is
\beq
\d{M}=\left(\d{X}\right)_{11}\,,
\hspace{15mm}
\d\mu=\left(\d{X}\right)_{22}\,.
\label{eq:Mmufixing}
\eeq

\subsection{Renormalization of the neutralino mass matrix
\label{subsec:neutrmass}}

The neutralino mass matrix at tree--level has the following
well--known form in the interaction (bino--wino--higgsino) basis:
\bea
        Y = \left(\begin{array}{cccc}
M' & 0 & - m_Z \sin\theta_W \cos\beta & m_Z \sin\theta_W
\sin\beta\\ 0 & M & m_Z \cos\theta_W \cos\beta & -m_Z \cos\theta_W
\sin\beta\\ - m_Z \sin\theta_W \cos\beta & m_Z \cos\theta_W
\cos\beta & 0 & -\mu\\ m_Z \sin\theta_W \sin\beta & -m_Z
\cos\theta_W \sin\beta & -\mu & 0
\end{array}\right) \,
 \label{eq:Ydef}
\eea
Since we assume CP conservation this matrix is real and symmetric.
It is diagonalized by the real matrix $N$.
\bea
N\,Y\,N^T = \left(\begin{array}{cccc}
\mnt{1} & 0 & 0 & 0 \\
0 & \mnt{2}  & 0 & 0 \\
0 & 0 & \mnt{3}  & 0 \\
0 & 0 & 0 & \mnt{4}  \\
\end{array}\right) \,.
 \label{eq:NYNT}
\eea
In analogy to the chargino case, the shift of $N$ is
\beq
\d N = \frac14\,
\left[\d Z^{\ti\chi^0}-\left(\d Z^{\ti\chi^0}\right)^T\right]\,
N\,.
\label{eq:dN}
\eeq
Note that $\d Z^{\ti\chi^0}_L=\d Z^{\ti\chi^0}_R$ due to the
Majorana nature.

The renormalization of $M'$ is fixed by
\beq
\d{M'}=\left(\d{Y}\right)_{11}=\d\left(N^TY_DN\right)_{11}=
\sum_{j=1}^4\e_j\left[\d\mnt{j}\left(N_{j1}\right)^2+
2\mnt{j}N_{j1}\,\d{N_{j1}}\right]\,,
\label{eq:Mpfixing}
\eeq
where $\e_j$ is the sign of $\mnt{j}$.

\subsection{Chargino/neutralino mass matrix at one--loop
\label{subsec:mmat1loop}}

In the (scale dependent) \drbar scheme the chargino/neutralino
mass matrix at one--loop level was already calculated some time ago
in~\cite{pierceetal}. Here we calculate it in the on--shell scheme.

Let us begin with the chargino mass matrix. One has to distinguish between
three types of the mass matrix: $X^0$ is the 'bare' mass matrix (or \drbar
running tree--level matrix), $X_{\rm tree}$ is the tree--level mass matrix
Eq.~(\ref{eq:chargmat}) in terms of the on--shell input parameters $M$, $\mu$,
$m_W$, $\tb$, and $X$ is the one--loop corrected mass matrix. We then have
the relations, on the one hand,
$$
X^0\,=\,X_{\rm tree}+\d_pX\,,
$$
where $\d_p$ means the variation of the parameters, and on the other hand,
$$
X^0\,=\,X+\d X\,.
$$
By eliminating $X^0$, one then gets
\beq
X\,=\,X_{\rm tree}+\d_pX-\d X\,=\,X_{\rm tree}+\D{X}\,,
\label{eq:Xsep}
\eeq
where $\D{X}$ is a finite shift.

We have already fixed $M$ and $\mu$. $m_W$ is fixed at the
physical (pole) W-mass. Concerning $\tb$, we follow the on--shell
fixing condition by~\cite{chankowski,dabelstein}
\beq {\rm Im}\left\{\hat{\Pi}_{A^0Z^0}(m_{A^0}^2)\right\}=0\,,
\label{eq:tbfixing}
\eeq
where $\hat{\Pi}_{A^0Z^0}(m_{A^0}^2)$ is the renormalized
self--energy for the mixing of the pseudoscalar Higgs boson $A^0$
and the $Z$ boson. This leads to the counter term
\begin{displaymath}
\frac{\d\tb}{\tb}~=~ \frac1{m_Z\sin{2\b}}\,{\rm Im}{\Pi}_{A^0Z^0}(m_{A^0}^2)\,.
\end{displaymath}
The gauge dependence of the fixing of $\tb$ and other fixing
conditions are discussed in~\cite{Youichi2} as well as in the parallel
sessions.

One obtains for the one--loop corrections $\D{X}$
\bea
 \D{X}_{11}&=&0\non\\[2mm]
 \Delta X_{12} &=& \left(\frac{\delta m_W}{m_W} + \cos^2\beta\,
\frac{\delta \tan\beta}{\tan\beta} \right)\,
 X_{12} - \delta X_{12}\,\label{eq:DX} \\[2mm]
 \non
 \Delta X_{21} &=& \left(\frac{\delta m_W}{m_W} - \sin^2\beta\,
\frac{\delta \tan\beta}{\tan\beta} \right)\,
 X_{21} - \delta X_{21}\,\\[2mm]\non
 \Delta X_{22} &=& 0\, ,
\eea

For the one--loop corrected neutralino mass matrix $Y$, we have analogously
\beq
Y\,=\,Y_{\rm tree}+\d_pY-\d Y\,=\,Y_{\rm tree}+\D{Y}\,,
\label{eq:Ysep}
\eeq
where $\d_p$ again means the variation of the parameters. With the
fixing of $M$ and $\mu$ in Eq.~(\ref{eq:Mmufixing}), $M'$ in
Eq.~(\ref{eq:Mpfixing}), and $\tb$ in Eq.~(\ref{eq:tbfixing}) one
gets
\begin{eqnarray}
  \non
\Delta Y_{11} &=& 0\,\\[2mm]
   \non
\Delta Y_{12} &=& - \delta Y_{12}\,\\[2mm]
  \non
\Delta Y_{13} &=& \left(\frac{\delta m_Z}{m_Z}+ \frac{\delta
 \sW}{\sW} - \sin^2\beta\, \frac{\delta \tan\beta}{\tan\beta}
 \right)\, Y_{13} - \delta Y_{13}\,\\[2mm]
  \non
\Delta Y_{14} &=& \left(\frac{\delta m_Z}{m_Z}+ \frac{\delta
 \sW}{\sW} + \cos^2\beta\, \frac{\delta \tan\beta}{\tan\beta}
 \right)\, Y_{14} - \delta Y_{14}\,\\[2mm]
  \non
\Delta Y_{22} &=& \delta M - \delta Y_{22}\, \,\\[2mm]
  \label{eq:DY}
\Delta Y_{23} &=& \left(\frac{\delta m_Z}{m_Z} - \tan^2\theta_W\,
\frac{\delta \sW}{\sW} - \sin^2\beta\, \frac{\delta
\tan\beta}{\tan\beta}
 \right)\, Y_{23} - \delta Y_{23}\,\\[2mm]
  \non
\Delta Y_{24} &=& \left(\frac{\delta m_Z}{m_Z} - \tan^2\theta_W\,
\frac{\delta \sW}{\sW} + \cos^2\beta\, \frac{\delta
\tan\beta}{\tan\beta}
 \right)\, Y_{24} - \delta Y_{24}\,\\[2mm]
  \non
\Delta Y_{33} &=&  - \delta Y_{33}\,\\[2mm]
  \non
\Delta Y_{34} &=& -\delta\mu - \delta Y_{34}\,\\[2mm]
  \non
\Delta Y_{44} &=& - \delta Y_{44}\, ,
\end{eqnarray}
Notice that $Y_{12}=Y_{21}$, $Y_{33}$ and $Y_{44}$ are no
more zero at one--loop level.

We could also have determined the on--shell values of $M$ and $\mu$
from the neutralino sector instead of the chargino sector by
$\D{Y}_{22}=\D{Y}_{34}=\D{Y}_{43}=0$, see Eq.~(\ref{eq:Ydef}) and
(\ref{eq:Ysep}). This would then imply corrections in the chargino
sector $\D{X}_{11}$ and $\D{X}_{22}$ different from zero.

By diagonalizing $X$ and $Y$, one gets the one--loop corrected chargino
and neutralino masses.

In practice, the chargino masses $m_{\ti{\x}^+_{1,2}}$ may be
known from experiment (e.g.~from a threshold scan). Then one first
calculates the tree--level parameters $M_{\rm tree}$ and $\mu_{\rm
tree}$ from $X_{\rm tree}$ (together with information on chargino
couplings to get a unique result). Then one calculates
$\d{X}_{ij}$ by Eq.~(\ref{eq:dX}) with $U$ and $V$ as the
tree--level matrices and then uses Eq.~(\ref{eq:DX}) to obtain
$\D{X}_{ij}$. The error that one starts from $M_{\rm tree}$ and
$\mu_{\rm tree}$ is of higher order. One proceeds in an analogous
way in the neutralino sector.

\begin{figure}[h!]
 \begin{center}
  \vspace*{-4mm}
 \mbox{\resizebox{100mm}{!}{\includegraphics{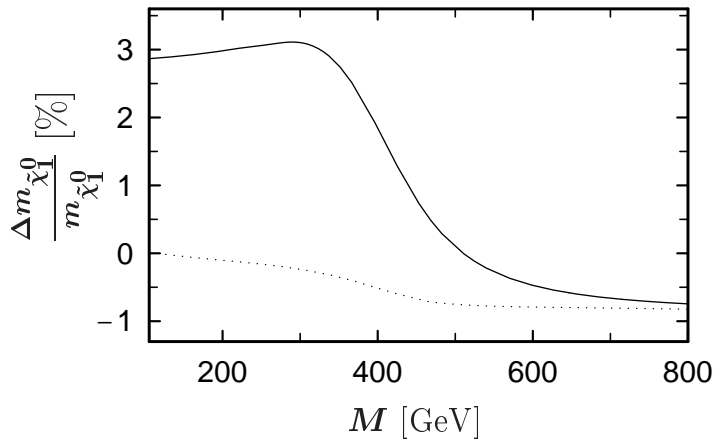}}}
 \vspace{-5mm}
 \caption[fig1]
 {\footnotesize Comparison of relative corrections to $\mnt{1}$ 
 \cite{chmasscorr}.
 The full line shows the case where the SUSY
 SU(5) GUT relation is assumed for the \drbar parameters $M$ and $M'$,
 and the on--shell $M'$ is determined from $M$ by the same relation.
 The dotted line corresponds to the case where the on--shell $M'$
 is an independent parameter but satisfies the SUSY GUT relation.
 Other parameters are
$\tan{\beta}=7$, $\{M_{\tilde Q_1},\,M_{\tilde Q},\,A\}=
\{300,300,-500\}$~GeV, and $\mu = -220$~GeV.}
 \label{fig:1}
 \end{center}
 \vspace{-7mm}
 \end{figure}

So far we have treated $M'$ as an independent parameter. The
situation is different if there is an intrinsic relation between
$M'$ and $M$ as, for instance, in SU(5) GUT models,
$M'=\frac{5}{3}\tan^2\theta_WM$ with $M'$ and $M$ as \drbar
parameters. If the same relation should hold for the on--shell
parameters, one has
\beq
 \Delta Y_{11}
 \;=\;\bigg(\frac2{\cos^2{\theta_W}}\,\frac{\delta\sW}\sW \,+\,
 \frac{\delta M}{M}\bigg)\,Y_{11}\;-\;\delta Y_{11}\,,
 \label{eq:DX11}
\eeq
instead of zero as in Eq.~(\ref{eq:DY}). The effect is shown in
Fig.~\ref{fig:1}.

Fig.~\ref{fig:1} shows the correction to $\mnt{1}$
as a function
of $M$ for $\tan{\beta}=7$, $\{M_{\tilde Q_1},\,M_{\tilde
Q},\,A,\,\mu\}= \{300,300,-500,-110\}$~GeV. ($M_{\tilde Q_1}$ and
$M_{\tilde Q}$ are soft breaking parameters of the first and third
squark generation, respectively, and $A$ is the trilinear soft breaking
parameter.) It can be seen that the difference between the case where the
GUT relation is assumed also for the on--shell $M'$ and $M$ and the
case where $M'$ is an independent parameter ($\D{Y}_{11}=0$)
with the value $0.498$ (as it would have by the GUT relation) is substantial.

By the way, Eq.~(\ref{eq:DX11}) would also be valid in the
anomalously mediated SUSY breaking model~\cite{AMSB,feng} where
$M'=11\tan^2\theta_WM$.

In conclusion, the method developed here is well suited for
extracting and studying the fundamental SUSY parameters $M$, $M'$,
$\mu$, etc. at higher order.

\section{Other on--shell methods of renormalizing \\ the
chargino/neutralino system \label{sec:othermethod}}

A different method for the on--shell renormalization of the
chargino and neutralino system was presented
in~\cite{Fritzshe-Hollik}. The same method was also used
in~\cite{Guasch-Hollik,Blank-Hollik,blank-hollik,freitas}.

Concerning the charginos, the input parameters are as before the
physical chargino masses $\mcha{1}$ and $\mcha{2}$. The parameters
$M$ and $\mu$ are calculated from the tree--level form of the mass
matrix, Eq.~(\ref{eq:chargmat}), by diagonalization,
Eq.~(\ref{eq:UXV}).

The on--shell condition is that the renormalized two--point function
matrix $\hat{\G}^{(2)}_{ij}$ get diagonal for on--shell external
momenta, i.e.
\beq
 {\rm Re}\,\hat{\G}^{(2)}_{ij}(p)\,u_j(p)=0\,,
 \label{eq:conditionGamma}
\eeq
for $p^2=\mcha{j}^2$. This fixes the non-diagonal elements of the
field-renormalization matrix. Its diagonal elements are
determined by normalizing the residues of the
propagators. These conditions then fix $\d{M}$ and $\d\mu$, but
they are different from the expressions Eqs.~(\ref{eq:Mmufixing}).

In the neutralino sector, one has the additional parameter $M'$
which is fixed together with its counter term $\d{M'}$ by a
neutralino mass, say $\mnt{1}$, and the appropriate on--shell
condition in analogy to Eq.~(\ref{eq:conditionGamma}). Hence one has
$$
  N\,Y\,N^T\,=\,
  {\rm diag}\left(m_1,m_2,m_3,m_4\right)\,=\,M_D\,,
  \hspace{20mm}
  {\rm with}\;
  m_1=\mnt{1}\;\;
  {\rm (pole\;mass)}
$$
$N$ is the mixing matrix in Eq.~(\ref{eq:NYNT}). $m_2$, $m_3$, $m_4$ are,
however, not yet the one--loop corrected pole masses $\mnt{i}$, $i=2,3,4$.
One has to find the momenta $p_i^2 = m^2_{\tilde\chi_i^0}$ so that
$$
{\mathrm Re}\left[\hat{\G}^{(2)}_{ii}(p_i)\right]\,u(p_i)\,=\,0\,,
$$
with
$$
\hat{\G}^{(2)}_{ij}(p_i)\,=\,
\left(\psla-m_i\right)\d_{ij}+\hat{\Sigma}_{ij}(p)\,,
$$
where $\hat{\G}^{(2)}$ is the renormalized two--point vertex function
and $\hat{\Sigma}$ is the renormalized self--energy.

It should be clear from above that the parameters $M$, $M'$, and
$\mu$ derived by this method are different from those in
section~\ref{subsec:chargmass} and~\ref{subsec:neutrmass}, and
in~\cite{chmasscorr}, and of course also from those in the \drbar
scheme~\cite{pierceetal}. They are ``effective'' parameters. However, the
observables (cross--section, branching ratios, particle masses,
\ldots) are the same in both methods.

\section{One--loop corrections to SUSY processes \label{sec:corr2proc}}

Let us start with a discussion of chargino production $$
e^+e^-\;\rightarrow\;\cha{i}\,\cham{j}\,. $$ Using the purely
on--shell renormalization scheme described above in
section~\ref{sec:othermethod}, the full one--loop corrections to
this process (including polarized beams) were calculated
in~\cite{blank-hollik}. This calculation is extremely cumbersome
as one has to compute a large number of graphs (self--energy
graphs, vertex corrections, box graphs) with all the particles of
the MSSM running in the loops. A further subtlety is due to the contributions
with virtual photons attached to two external charged particles.
To obtain an infrared finite result, real photon bremsstrahlung
from the external particles has to be taken into
account. Without the virtual photons the result would not be UV
finite. They are required to cancel the divergence coming from the
photino component of the virtual neutralinos.

\begin{figure}[h!]
 \begin{center}
 \mbox{\resizebox{150mm}{!}{\includegraphics{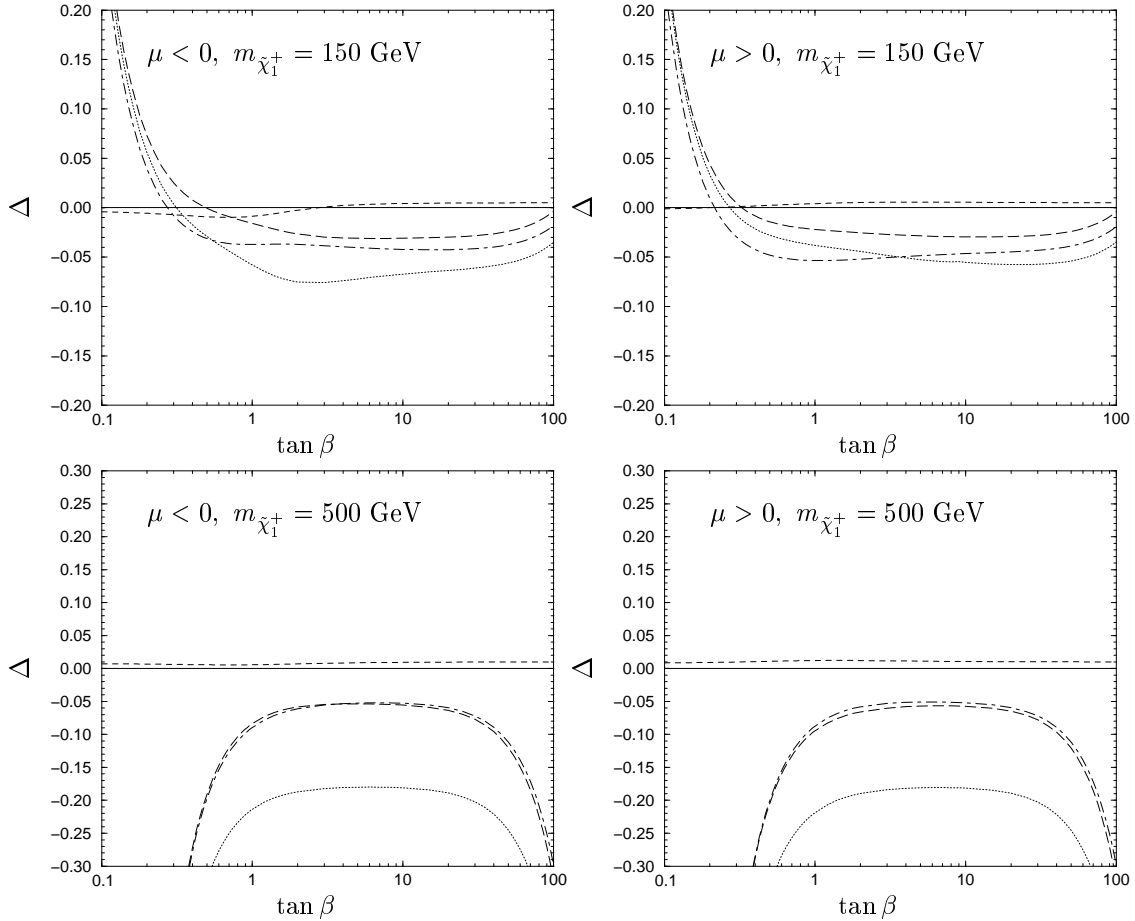}}}
 \caption[fig2]
 {\footnotesize One--loop corrections to the cross--section of
 $e^+ e^- \rightarrow \tilde\chi_1^+ \tilde\chi_1^-$, normalized
 to the Born approximation \cite{blank-hollik}. Upper (lower) row:
 $M_2 = 200$~GeV (800~GeV), $\sqrt{s} = 500$~GeV (1~TeV),
 $M_{\rm SUSY} = 500$~GeV. The various lines show the contributions
 from different subsets of diagrams: the dash--dotted, long dashed,
 dashed and dotted line corresponds to (s)top/(s)bottom loops,
 (s)fermion loops, SM fermion loops and all loops without ISR.}
 \label{fig:hollik}
 \end{center}
 \end{figure}

Fig.~\ref{fig:hollik} shows the relative one--loop correction
$\D=\frac{\s-\s_0}{\s}$ ($\s_0$ being an improved Born cross--section)
to the total cross--section of
$e^+e^-\;\rightarrow\;\cha{1}\,\cham{1}$ as a function of $\tb$ at
$\sqrt{s}=500$~GeV and $\sqrt{s}=1$~TeV, with the parameters as
indicated in the figure caption (The initial state radiation (ISR)
is separated off). Typically, the corrections are between $5$ and
$10$\%, but can go up to more than 20\% at $\sqrt{s}=1$~TeV. The
figure also exhibits the contributions from various subsets of
graphs. One clearly sees that the (s)top/(s)bottom loops are by
far not enough to explain the full correction. The figure
demonstrates the necessity of a complete full one--loop
calculation.

A complete one--loop calculation for chargino production in
$e^+e^-$ annihilation was also performed in~\cite{Ross}. However,
the renormalization scheme is different. For the charginos as
external particles the subtractions are made on--shell. The masses
of all particles are also taken to be physical, but all other
parameters (as couplings) are considered to be in the \drbar
scheme at the scale $m_Z$. In~\cite{Ross} the one--loop corrections
to the \xp production helicity amplitudes (with polarized beams)
are calculated. However, pure QED corrections involving loops of
photons were omitted.

\begin{figure}[h!]
 \begin{center}
 \mbox{\resizebox{150mm}{!}{\includegraphics{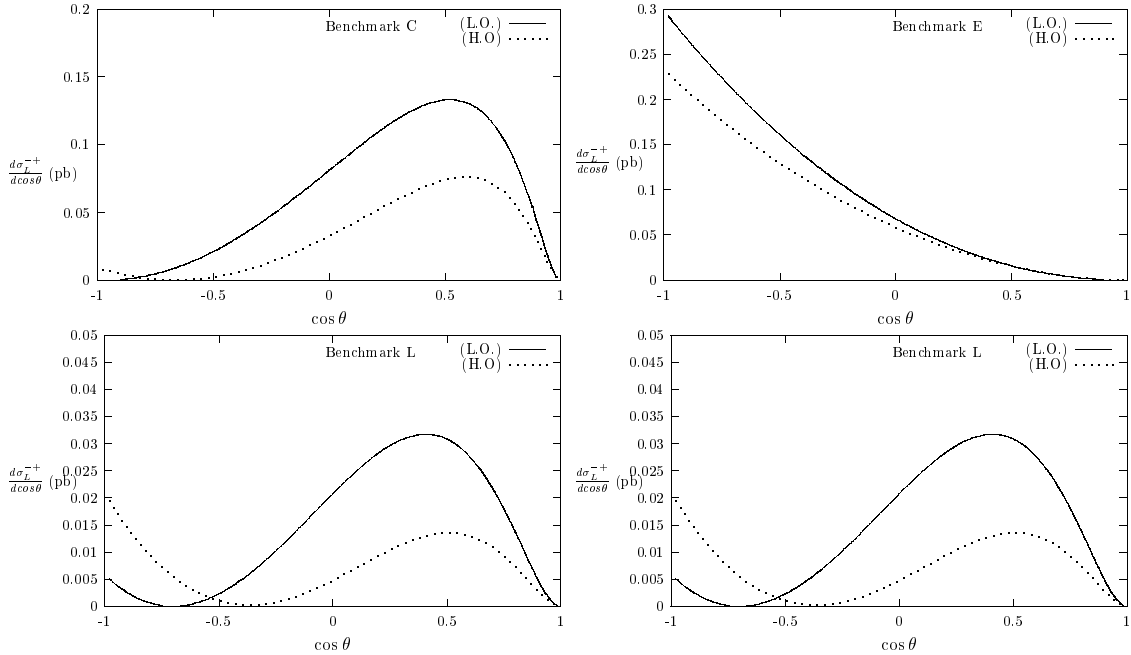}}}
 \caption[fig3]
 {\footnotesize Lowest and higher order cross--section $\sigma_L^{-+}$ for
 $e^+ e^- \rightarrow \tilde\chi_1^+ \tilde\chi_1^-$ with left--polarized
 electrons and negative (positive) helicity of the chargino
 $\tilde\chi_1^+$ ($\tilde\chi_1^-$) \cite{Ross}.}
 \label{fig:ross}
 \end{center}
 \end{figure}

In Fig.~\ref{fig:ross} $\frac{d\s^{-+}_L}{d{\cos{\theta}}}$ is shown as a
function of $\cos{\theta}$ for various benchmark
scenarios~\cite{battaglia}. $\s^{-+}_L$ is the cross--section for
producing a negative helicity chargino and a positive helicity
anti-chargino with a left--handed electron. Shown are the tree--level
cross--sections and the one--loop corrected ones. The corrections can
be very large ($>50\%$), especially if the cross--section is low.

The other processes for which full one--loop corrections were
calculated~\cite{freitas} are
\bea
 \non
 e^+e^-&\rightarrow&
 \ti{\mu}_R^+ \ti{\mu}_R^-\,\;,
 \ti{\mu}_L^+ \ti{\mu}_L^-\,,
 \\
 e^\pm {e}^-&\rightarrow&
 \ti{e}^\pm_R\ti{e}^-_R\,\;,
 \ti{e}^\pm_L\ti{e}^-_L\,\;,
 \ti{e}^\pm_R\ti{e}^-_L\,\;,
 \ti{e}^\pm_L\ti{e}^-_R\,\;,
\eea
with polarized $e^+$ and $e^-$ beams.

Especially for $e^+e^-\rightarrow\ti{e}^+\ti{e}^-$ the calculation
is rather complex because of the \xz exchanges in the
t-channel. One has to \ren the \xz sector.

\begin{figure}[h!]
 \begin{center}
 \hspace{-7mm}
 \mbox{\resizebox{80mm}{!}{\includegraphics{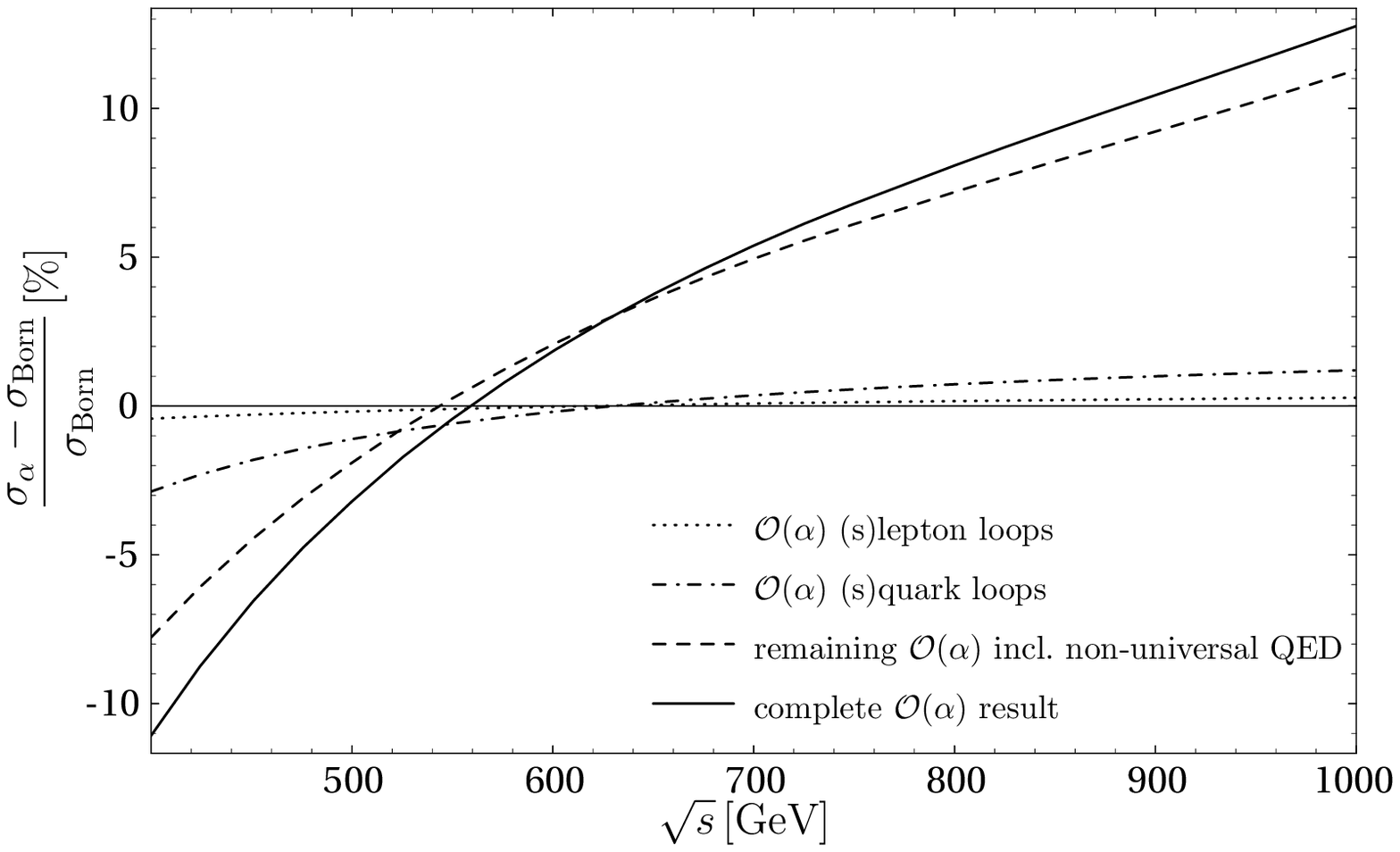}}}
 \hspace{-0mm}
 \mbox{\resizebox{80mm}{!}{\includegraphics{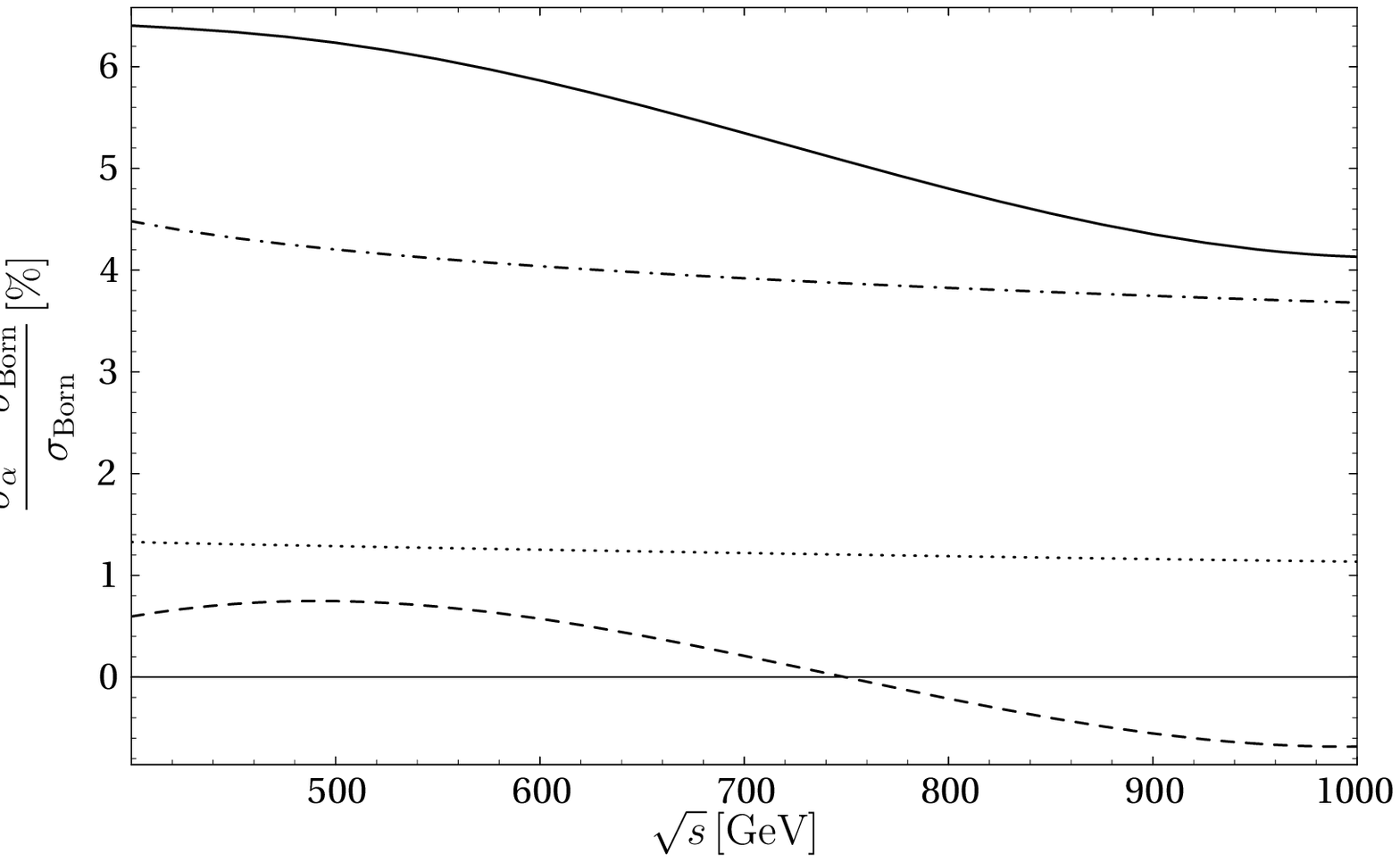}}}
 \hspace{0mm}
 \caption[fig4]
 {\footnotesize Electroweak corrections to the cross--section of (a)
 $e^+ e^- \rightarrow \tilde e_R^+ \tilde e_R^-$, (b)
 $e^- e^- \rightarrow \tilde e_R^- \tilde e_R^-$, relative to the
 Born cross--section \cite{freitas}. Input parameters correspond to the
 SPS1 scenario.}
 \label{fig:freitas}
 \end{center}
 \end{figure}

Fig.~\ref{fig:freitas}a shows the $\sqrt{s}$ dependence of the full
electroweak corrections ${(\s-\s_{\rm Born})}/{\s_{\rm Born}}$ for
$e^+e^-\rightarrow\ti{e}_R^+\ti{e}_R^-$. The input values
correspond to the SP1 scenario~\cite{snowmass}, with the mSUGRA
point $m_0=100$~GeV, $M_{1/2}=250$~GeV, $A_0=-100$~GeV, $\tb=10$,
$\mu>0$. The ``universal'' initial state radiation (ISR) has been
subtracted. However, the process dependent QED corrections are
included. For illustration, also the contributions from various
subsets of diagrams are shown. One can see that the (s)lepton
loops and (s)quark loops do not account for the whole effect. This
means that the gauge boson, Higgs boson, gaugino and higgsino
exchanges are important. In total, the corrections are
$\leq|10|\%$. Fig.~\ref{fig:freitas}b shows the corresponding electroweak
correction as a function of $\sqrt{s}$ for
\\$e^-e^-\rightarrow\ti{e}_R^-\ti{e}_R^-$.

\begin{figure}[h!]
 \begin{center}
 \hspace{-7mm}
 \mbox{\resizebox{80mm}{!}{\includegraphics{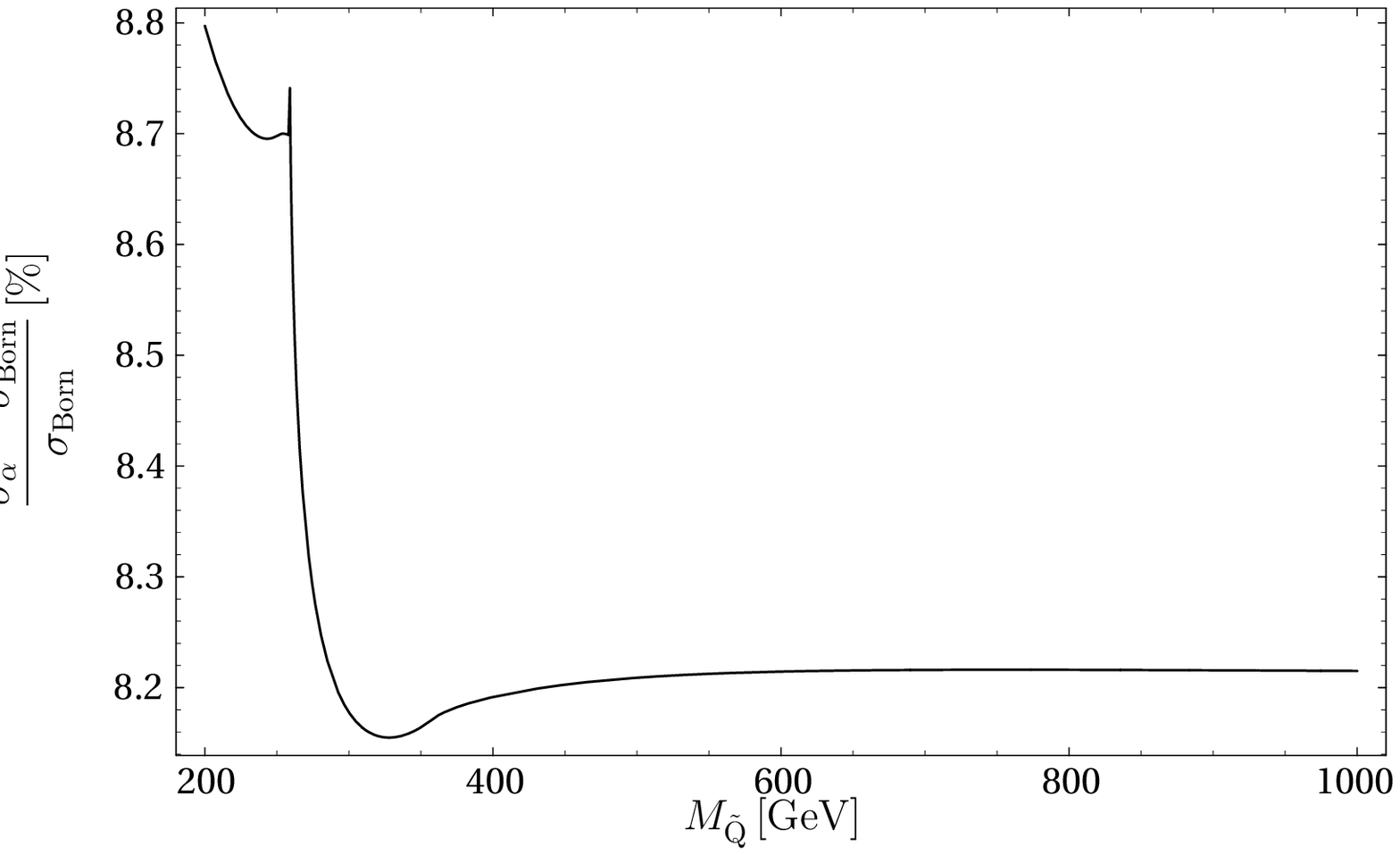}}}
 \hspace{-0mm}
 \mbox{\resizebox{80mm}{!}{\includegraphics{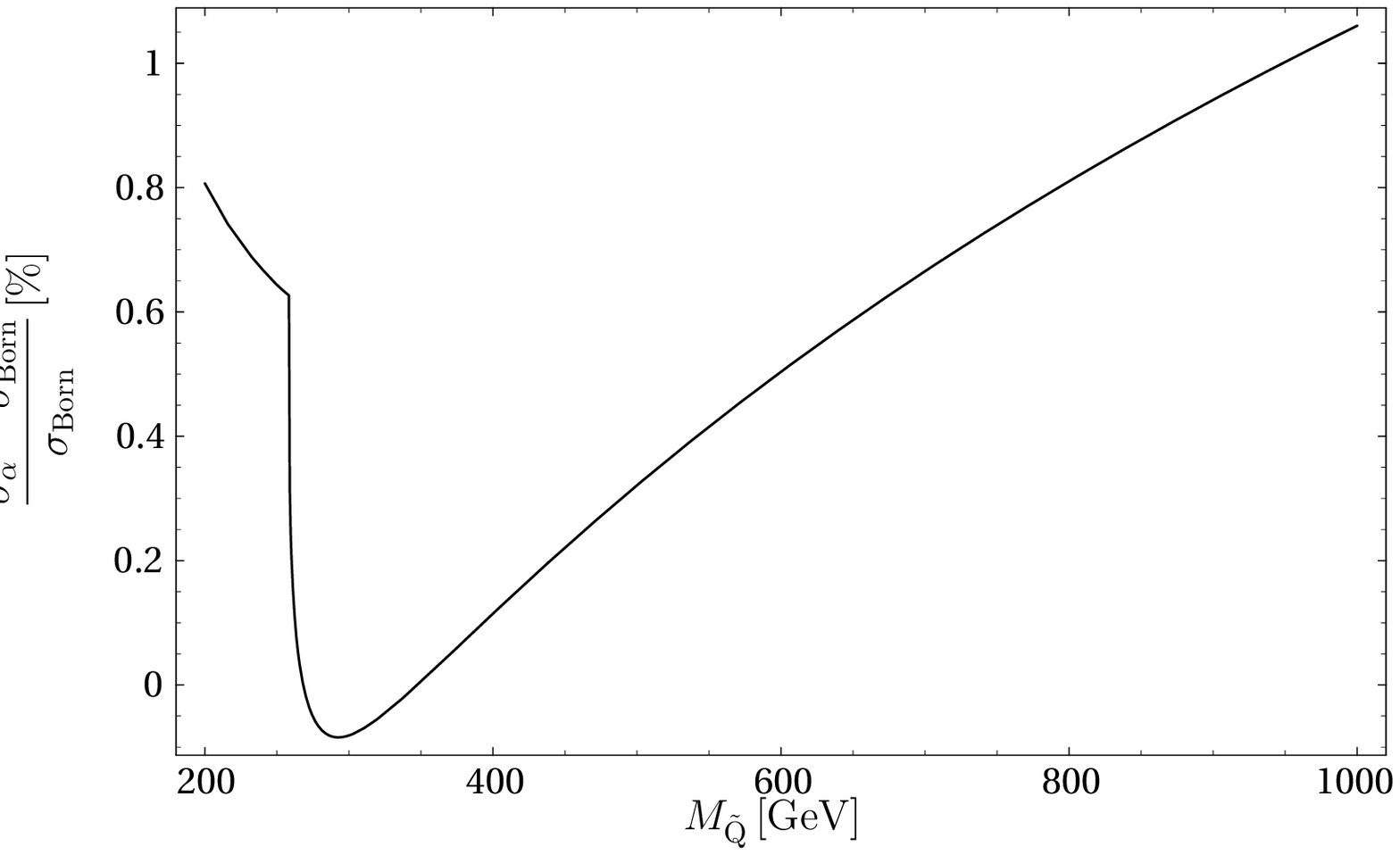}}}
 \hspace{0mm}
 \caption[fig5]
 {\footnotesize Dependence of the relative one--loop corrections
 $\Delta_\alpha$ on the universal squark soft breaking parameter
 $M_{\tilde Q}$ to the cross--section of (a) $e^+ e^- \rightarrow
 \tilde\mu^+_R \tilde\mu^-_R$ for $\sqrt{s} = 500$~GeV, and (b)
 to $e^- e^- \rightarrow \tilde e^-_R \tilde e^-_R$ for $\sqrt{s}
 = 400$~GeV \cite{freitas}. Input parameters correspond to the SPS1 scenario.}
 \label{fig:freitas2}
 \end{center}
 \end{figure}

In Fig.~\ref{fig:freitas2} the dependence of the one--loop
corrections is exhibited as a function of the soft breaking
parameter $M_{\ti{Q}}=m_{\ti{q}_L}=m_{\ti{u}_R}=m_{\ti{d}_R}$.
(The other parameters are as in the SPS1 scenario.)
Fig.~\ref{fig:freitas2}a shows the behaviour for $\ti{\mu}_R^+
\ti{\mu}_R^-$ production at $\sqrt{s}=500$~GeV. One sees that for
large $M_{\ti{Q}}$ the correction reach an asymptotic value being
an indication for the decoupling of the squarks. In contrast,
Fig.~\ref{fig:freitas2}b shows the corresponding
$M_{\ti{Q}}$-dependence for $\ti{e}_R^-\ti{e}_R^-$ production at
$\sqrt{s}=400$~GeV. In this case, the size of the radiative
corrections increases with $M_{\ti{Q}}$ showing no decoupling. This
is due to the so--called superoblique corrections~\cite{cheng}.
These stem from squark--quark loops in \xz self--energies.

For the squark pair production process
$$
e^+e^-\rightarrow\;\;\ti{q}_i \;\bar{\ti{q}}_j\,,
\hspace{20mm}
{\rm with}\;\;
j=1,2\;;\;\;
\ti{q}=\ti{t},\ti{b}
$$
only the SUSY--QCD corrections \cite{bartl} and the Yukawa
corrections due to quark and squark loops have been
calculated~\cite{21a}. Whereas the QCD corrections are typically
15--20\%, the Yukawa corrections are~$<|10\%|$ of the tree--level
cross--section.

Very recently, the next--to--leading SUSY--QCD corrections have been
calculated~\cite{brandenburg} for
\bea
 \non
 e^+e^-\rightarrow\;\;q\;\bar{q}\;g
 \\
  \non
 e^+e^-\rightarrow\;\;\ti{q}\;\bar{\ti{q}}\;g
 \\
  \non
 e^+e^-\rightarrow\;\;q\;\bar{\ti{q}}\;\ti{g}
\eea where $q$ denotes a light quark, and the squarks have no
mixing. By comparing these reactions the equality of the
$q\,q\,g$, the $\ti{q}\,\ti{q}\,g$, and the Yukawa coupling
$q\,\ti{q}\,\ti{g}$ in the supersymmetric limit ($m_{\ti{q}}=m_q$,
$m_{\ti{g}}=m_g=0$) can be tested.

\begin{figure}[h!]
 \begin{center}
 \mbox{\resizebox{80mm}{!}{\includegraphics{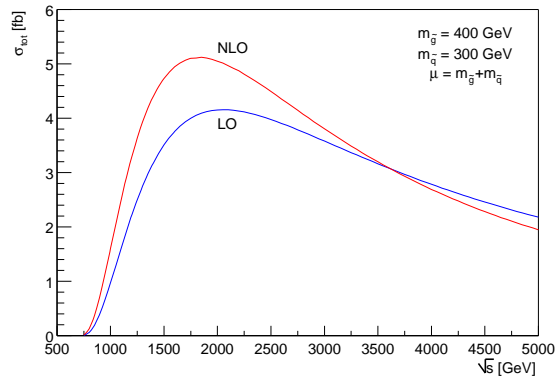}}}
 \hspace{0mm}
 \caption[fig7]
 {\footnotesize Total cross--section for $e^+ e^- \rightarrow
 \tilde q\, \bar q\, \tilde g\, (g) + q\ \bar{\!\tilde q} \, \tilde g\, (g)$
 in leading (LO) and next--to--leading (NLO) order, as a function of
 $\sqrt{s}$~\cite{brandenburg}.}
 \label{fig:brandenburg}
 \end{center}
 \end{figure}

Fig.~\ref{fig:brandenburg} shows as a function of $\sqrt{s}$ the
cross--section in leading order (LO) and next--to--leading order(NLO)
for
$e^+e^-\rightarrow\,\;\ti{q}\,\bar{q}\,\ti{g}\,(g)\,+\,q\ \bar{\!\ti{q}}\,
\ti{g}\,(g)$,
summed over $u$, $d$, $c$, $s$, $b$ quarks; $m_{\sg}=400$~GeV,
$m_\sq=300$~GeV and $\mu=m_\sg+m_\sq$. The cross--section goes up
to 5~fb. At the peak the corrections are about ~20\%, enhancing
the LO cross--section.

\subsection{Decays of supersymmetric particles \label{subsec:decays}}

The SUSY--QCD corrections to the decays
\bea
\non
\sq_i&\rightarrow&q\,\neu{i}\;, \hspace{30mm} i=1,..,4\\
\non
\sq_i&\rightarrow&q\,\chapm{k}, \hspace{30mm} k=1,2
\eea
were already calculated some time ago~\cite{beenaker}. The full
electroweak corrections were computed recently in~\cite{Guasch-Hollik}.

\begin{figure}[h!]
 \begin{center}
 \mbox{\resizebox{75mm}{!}{\includegraphics{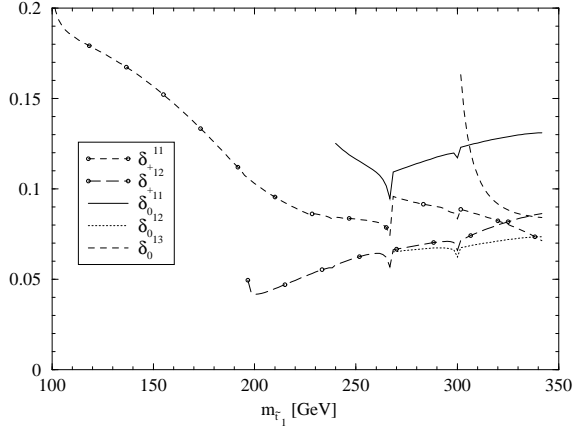}}}
 \hspace{0mm}
 \caption[fig8]
 {\footnotesize Relative electroweak radiative corrections to the partial
 widths of the lightest stop decaying into charginos and
 neutralinos~\cite{Guasch-Hollik}.
 $\delta_+^{11}$ corresponds to $\tilde t_1 \rightarrow b\, \tilde\chi_1^+$,
 $\delta_+^{12}$ to $\tilde t_1 \rightarrow b\, \tilde\chi_2^+$,
 $\delta_0^{11}$ to $\tilde t_1 \rightarrow t\, \tilde\chi_1^0$,
 $\delta_0^{12}$ to $\tilde t_1 \rightarrow t\, \tilde\chi_2^0$ and
 $\delta_0^{13}$ to $\tilde t_1 \rightarrow t\, \tilde\chi_3^0$.}
 \label{fig:guasch}
 \end{center}
 \end{figure}

Fig.~\ref{fig:guasch} shows the $m_{\st_1}$ dependence of the
relative electroweak corrections $\d$ to the partial widths of
$\st_1\rightarrow{b}\,\cha{1},\,{b}\,\cha{2}$,
$\st_1\rightarrow{t}\,\cha{1},\,{t}\,\neu{2},\,{t}\,\neu{3}$ for
the parameter set\\
$\{M,\,\mu,\,\tb,\,\t_\st,\,m_{\sb_1},\,\t_\sb,\,M_{H^+}\}=
\{150\,{\rm GeV},-100\,{\rm GeV},\,4,\,0.6,\,300\,{\rm GeV},\,0.3,\,120\,{\rm
GeV}\}$. They go up to 20\%. The various spikes are due to the
opening of other decay channels. Although the QCD corrections are
usually dominant, the electroweak corrections can be of the same size
in certain regions of the parameter space. \vspace{1mm}

The decays of a squark into a quark and gluino, and of a gluino
into a squark and quark
\bea
\non
\sq_i&\rightarrow&q\,\sg \\
\sg&\rightarrow&\sq_i\,q
\hspace{30mm} i=1,2 \non
\eea
were calculated including one--loop SUSY--QCD corrections
in~\cite{zerwas}. The corrections can go up to 50\%.

The Yukawa corrections up to ${\cal O}(\a_{ew}m_q^2/m_W)$ to
$\st_2\rightarrow t\,\sg$ and
$\sg\rightarrow\bar{t}\,\st_1\;+c.c.$ were calculated
in~\cite{hongshang}. They reach values of $\sim10\%$.

The SUSY--QCD corrections to the decays
\beq
\st_i\rightarrow\sb_jW^+\,,\;\;\;\sb_i\rightarrow\st_jW^-
\hspace{10mm}{\rm \;and\;}\hspace{10mm}
\st_2\rightarrow\st_1Z^0\,,\;\;\;\sb_2\rightarrow\sb_1Z^0
\label{eq:stsbdecays}
\eeq
were computed in~\cite{susyqcdhiggs}. The squark decays into Higgs
bosons
\bea
\sq_i&\rightarrow&\sq_j^{(\prime)}\,H_m
\hspace{30mm} H_m=\{h^0,\,H^0,\,A^0,\,H^\pm\}
\label{eq:sqH}\,.
\eea
including SUSY--QCD corrections have been treated in~\cite{arhrib, bartlD59}.
For the Higgs boson decays into squarks
\bea
\non
H_m&\rightarrow&\sq_i\,\bar{\ti{q}}^{(\prime)}
\,,
\eea
these corrections were calculated in~\cite{arhrib,abartl}.
In the on--shell scheme, they can become very large for large $\tb$
(as in the case of $H_l\rightarrow{q}\bar{q}$), making the
perturbation expansion quite unreliable. An improvement of the
perturbation calculation was proposed in~\cite{eberl}. This
is achieved by using the SUSY--QCD running quark mass $m_q(Q)_{\rm
MSSM}$ and the running trilinear coupling $A_q(Q)$ in the
tree--level coupling. However, the mixing angle $\t_\sq$ is kept
on--shell in order to cancel the $\sq_1-\sq_2$ mixing squark
wave--function correction.

The one--loop corrected decay widths for
\bea
 \non
 H_m&\rightarrow&\neu{k}\,\neu{l}\\
 \neu{k}&\rightarrow&H_m\,\neu{l}
\eea
were calculated in~\cite{H0toNeu} taking into account all fermion
and sfermion loop contributions. The neutralino mass matrix was
renormalized as described in section~\ref{subsec:mmat1loop}.
\begin{figure}[h!]
 \begin{center}
 \hspace{-8mm}
 \mbox{\resizebox{84mm}{!}{\includegraphics{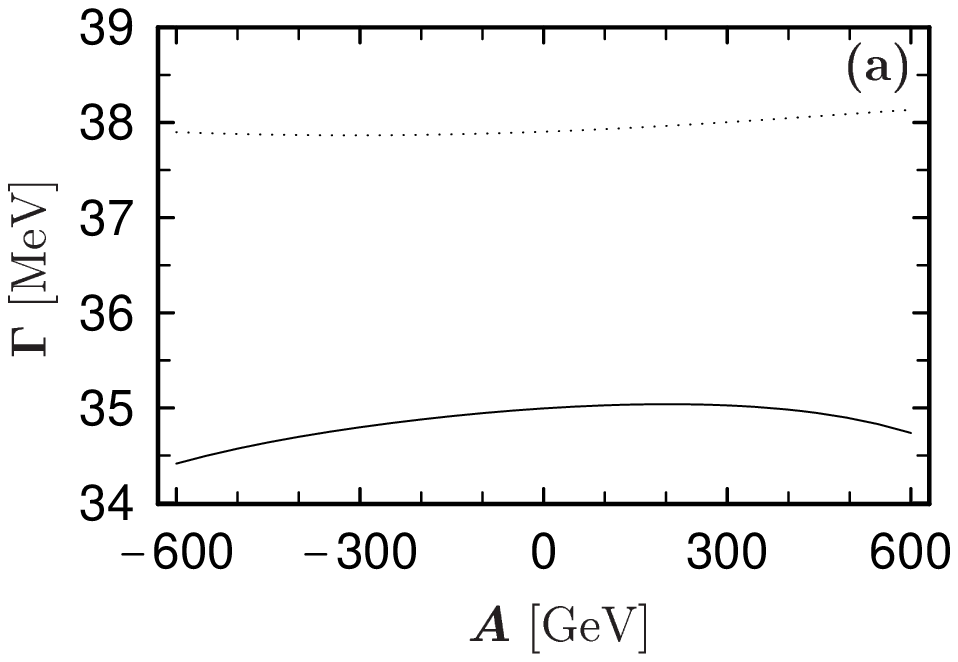}}}
 \hspace{-3mm}
 \mbox{\resizebox{84mm}{!}{\includegraphics{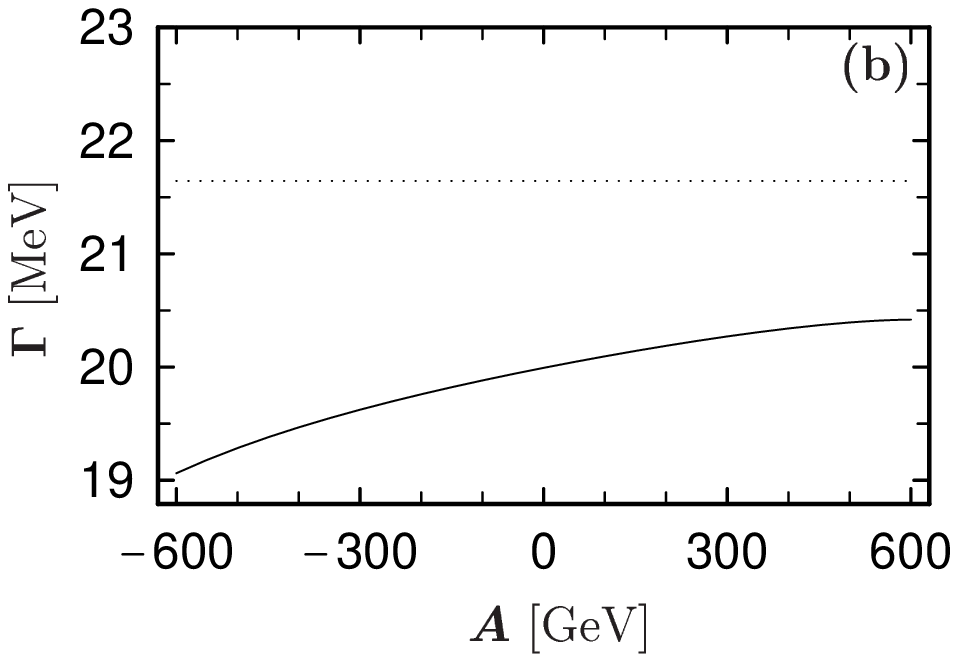}}}
 \hspace{-10mm}
 \vspace{-0mm}
 \caption[fig9]
{\footnotesize The widths of the decays $H^0\to\ch^0_1+\ch^0_2$~(a) and
$A^0\to\ch^0_1+\ch^0_3$~(b) as a function of $A$~\cite{H0toNeu}.
The dotted line corresponds to the tree--level width, the solid line
corresponds to the full correction. The parameters are
$\tan{\beta}=10$, $M=500$~GeV, and $\mu=150$~GeV~(a) and
$\tan{\beta}=50$ and $M=\mu=300$~GeV~(b).}
 \label{fig:ours}
 \end{center}
 \vspace{-5mm}
 \end{figure}

Fig.~\ref{fig:ours} shows the correction to the width of
$H^0\rightarrow\neu{1}\neu{2}$ and $A^0\rightarrow\neu{1}\neu{3}$
as a function of the trilinear coupling $A$ for the parameters
as given in the figure caption.

The decays into charginos
\bea
 \non
 H^0(A^0)&\rightarrow&\cha{i}\cham{j}\\
 H^+&\rightarrow&\cha{i}\neu{l}
 \non
\eea
were treated in the same approximation in~\cite{zhang}.


\section{Conclusions}

Precision experiments which will be possible at a future linear
$e^+ e^-$ collider will require equally precise theoretical calculations
of cross--sections, decay branching ratios and other observables,
including higher order corrections. For doing such calculations, we have
presented the on--shell renormalization of the sfermion and chargino/neutralino
system in the MSSM. We have worked out the appropriate renormalization
conditions, especially for the mixing matrices. We have discussed the
calculations of one--loop corrections to various SUSY processes: sfermion
and chargino/neutralino production in $e^+ e^-$--annihilation, the
two--body decays of sfermions and the decays of Higgs
bosons into SUSY particles. In a few cases the full electroweak corrections
have already been calculated. They clearly show that taking only a subset
of diagrams, for instance, only (s)top/(s)bottom loops, is not sufficient.
The electroweak corrections are typically between 5 and 15\%, but can go up
to larger values for certain parameters. Although the QCD corrections are
usually the largest ones, the electroweak corrections can be of the
same size in certain cases.

\section*{Acknowledgements}
The author is deeply indebted to his collaborators on this subject,
H.~Eberl, M.~Kincel and Y.~Yamada. He also thanks for discussions with
A.~Bartl, S.~Kraml, W.~Porod, V.~Spanos and C.~Weber. The work was
supported by the ''Fonds zur F\"orderung der wissenschaftlichen Forschung
of Austria'', project no. P13139-PHY.



\end{document}